\documentclass{article}

\usepackage[english]{babel}

\usepackage{amsmath}
\usepackage{graphicx}
\usepackage[colorlinks=true, allcolors=blue]{hyperref}

\title{Towards Automated Verification of Logarithmic Arithmetic}
\author{Mark G. Arnold, Thomas A. Bailey and John R. Cowles\footnote{This work was supported in part by Computational Logie, Inc. of Austin, TX. The views and conclusions contained in this paper are these of the authors and should not be interpreted as representing the official policies, either expressed or implied, of the University of Wyoming or Computational Logic, Inc.}\\
{\small University of Wyoming, Computer Science Dept. Box 3682, Laramie, WY 82071}}

\date{}

\begin{document}
\maketitle
\begin{abstract}
Correctness proofs for floating point programs are difficult to verify. To simplify the task, a similar, but less complex system, known as logarithmic arithmetic can be used, The Boyer-Moore Theorem Prover, NQTHM, mechanically verified the correctness of a simple implementation of logarithmic arithmetic. It also verified some useful theorems about accumulated relative error bounds for addition, multiplication and division in this logarithmic number system. These theorems were used to verify a program that approximates $e^x$ using a truncated Taylor series.
Axioms that characterize the finite precision of the logarithmic system using a rational base, $b$, were shown by the prover to be satisfiable for any choice of $1 < b < 2$. The prover verified the correctness of a function for converting an arbitrary rational value to a logarithmic representation. It also verified that multiplication and division implementations produce exact results for exact inputs, and that addition implementation produces a result as accurate as possible for exact inputs.
When these operations are used in combination by a program, such as evaluating a polynomial, the relative error increases in a way that can be bounded by simple expressions, referred to here as tolerances. Several mechanically verified theorems about tolerances allow us to construct mechanically verified proofs about logarithmic arithmetic programs. Although similar to interval arithmetic, tolerances are especially suited to logarithmic arithmetic.\end{abstract}

\section{Introduction}

It is ironic that floating point computation, which is seemingly one of the most rigorous and well defined areas of computer systems, should be one of the last areas to see application of automated formal techniques. Although languages [23], operating systems [6], and even hardware [12] have been mechanically verified, little has been done with floating point programs.

Recently, Wilding [36] has mechanically checked the proof of a simple floating point search program that calculates the mid-point in an interval, and proceeds recursively on either the left or right side until a root of a continuous function is found. Even though the only floating point computation in this program is
$(a+b)/2$, the proof is over fifty pages long. Wilding claims this is the {\it only} floating program whose correctness proof has ever been mechanically verified, although some proofs of floating point correctness have been manually constructed [11]. Higher order arithmetic algorithms [28] can be appended to conventional floating point programs to verily that a result for a particular input is accurate, but to our knowledge, no such floating point result verifying algorithm has had its proof of correctness mechanically verified. Although we have no reason to doubt the correctness of such algorithms, it would be desirable to have a verified implementation. Similarly, interval machine arithmetic [21] can be used to obtain upper and lower bounds for an entire computation, but this doubles the storage requirements, and often produces impractical bounds when applied naively.
Since mechanical verification of correct floating point programs appears to be a challenging problem, it is sensible to choose the simplest possible model af floating point. Several studies of floating point behavior [20, 3, 8] have used {\em logarithmic arithmetic} to model an idealized floating point system,
Logarithmic arithmetic maintains constant relative precision over the entire range of representable values [3], Because of this, the error analysis for logarithmic arithmetic is similar to, but somewhat easier than, the analysis for conventional floating point arithmetic, The ranges and relative precisions when the two arithmetics use the same word size are nearly identical, except the relative precision of conventional base two floating point representations wobble by a factor of two,
Logarithmic arithmetic is more than just a theoretical tool. It has been used in a variety of practical applications, including: battery-opernted hearing aids [24], circuit simulation [19], control algorithms [35], digital filters [29], Fast Fourier Transforms [32], graphics [16], image processing [15], matrix inversion [26], neural networks [1], speech recognition [25] and tomography [31]. Several successful implementations have been reported in the literature [14, 34, 37], Developers of logarithmic arithmetic software for personal computers [27, 10, 2] claim that on machines with no hardware support for floating point, programs with intensive real computations execute much faster with logarithmic arithmetic than with conventional software floating point arithmetic. The cost and power consumption of many math coprocessor chips makes them impractical for applications which demand low power consumption and/or low cost microprocessors that lack on chip floating point hardware (e.g., 68020, 86C010 or 80486SX), and so logarithmic arithmetic is attractive for such applications.
There has been moderate interest lately in the so called SLI number system, which offers freedom from overflow in most computations [33]. Should this unusual number system ever gain wider acceptance, there will be a need ta provide mechanically checked proofs of SLI correctness, just as such a need currently exists for floating point and logarithmic arithmetics, In fact, SLI, floating point, and logarithmic number systems are members of a class of number systems that are closely related. SLI is a generalization of logarithmic arithmetic, and SLI representations near one are identical, bit for bit, to their logarithmic counterparts. SLI differs only in how numbers that approach zero and infinity are represented, Therefore, our work here provides some insight not only in how mechanically checked proofs can be constructed for logarithmic arithmetic programs, but also how they might be done for floating point and SLI programs.

\section{Finite Precision and the Base}
The $\log_b$ function, for any real positive $b\neq 1$, maps positive reals onto the reals. If we could store logarithms with infinite precision, real arithmetic could be carried out exactly. Let $x$ and $y$ be positive real values represented by $l_x=\log_b(x)$ and $l_y = \log_b(y)$ respectively.
Then $x-y$, $x/y$, and $x + y$ are represented by the real (often irrational) representations: $l_x + l_y$, $l_x - l_y$, and $s_b(ly-l_x) + l_x$, where the transcendental function $s_b(z) = \log_b(b^z +1)$ is known as the addition logarithm.

Of course, to be practical in high speed, cost sensitive applications, the representations $l_x$ and $l_y$ can only be stored with a finite number of bits. Although how $l_x$ and $l_y$ are truncated is arbitrary, the simplest approach is to use the integers 
$X = \lfloor l_x \rfloor$  
as the representation of the real value $x$, and 
$Y = \lfloor l_y \rfloor$  
as the representation of $y$. 
In this case, the choice of $b$ is no longer arbitrary, since the relative precision of the logarithmic system, $F$, determines the base,
\begin{equation*}
    b=\sqrt[2^F]{2}.
\end{equation*}
Although such a representation is often thought of as a logarithm of arbitrary base (often two) stored as a fixed point number whose scaling determines the precision [14], the integer view of the representation is completely equivalent [20]. We choose this way of viewing the truncation, since it facilitates the formalization and machine verification described below.
Using the irrational given above for $b$ is as impractical for the purpose of automated verification as trying to store the irrational $l_x$. For a given precision $F$ and a finite dynamic range, there is a rational choice for $b$ that would produce the same results, bit for bit, as the irrational choice. This is because the irrational base may be approximated by a rational number as closely as needed to make all results identical within the finite range. Thus in the formalization of logarithmic arithmetic, the base $b$ is assumed to be a rational larger than 1.
When $b=\sqrt[2^F]{2}$, we note that $F = \log_2(\log_b 2)$. So for any choice of base, $b$, the precision is defined to be the integer, $F$, given by
\begin{equation*}
F = \lfloor \log_2(\log_b 2)\rfloor.    
\end{equation*}
The condition, $F \geq 0$, forces $b\leq 2$. Except for showing the axioms presented below are satisfiable, there is no interest in the cases when $F = 0$, so it is safe to exclude the possibility that $b = 2$.

\section{Axioms}

The base is characterized by $b= P/Q$, where $P$ and $Q$ are integers that satisfy $0<Q< P<2\cdot Q$. Since $P$ and $Q$ are integers, it is not possible for $Q = 1$, and so slightly stronger conditions can be imposed on $P$ and $Q$ as the first axiom in our formalization:
\begin{equation}
    1 < Q < P < 2\cdot Q.
\end{equation}
Given a base, $b = P/Q$, that satisfies (1), an integer valued function, $S(z)$, known as the quantized addition logarithm, is required to approximate the transcendental function $s_b(z) = \log_b(b^z + 1)$. The asymptotic properties of $s_b$, make it feasible to implement $S(z)$ with finite tables. 
For large positive inputs, $s_b(z)$ approaches z. 
For negative inputs with large absolute value, $s_b(z)$ approaches 0. Since the curve $y = s_b(z)$ approaches its asymptotes from above, when $S(z)$ is defined by
\begin{equation*}
S(z) = \lfloor s_b(z) \rfloor,    
\end{equation*}
then there is a nonnegative integer, SEZ, and a nonpositive integer, Sez, known as the essential zero, such that for all integers $Z$ larger than SEZ, $S(Z) = Z$ and for all integers $Z$ less than Sez, $S(Z) = 0$.

Since 
$S(-Z) < s_b(-Z) < S(-Z)+1$, 
we see that 
$S(-Z)+Z < s_b(-Z)+Z < S(-Z)+ Z +1$. 
Then 
$S(-Z) + Z < s_b(Z) < S(-Z) +Z+1$, 
since $s_b(z) = s_b(-z) + z$ is a well known property of the real valued addition logarithm [1]. Thus for all integers $Z$, a similar property applies to the quantized addition logarithm,
\begin{equation*}
S(Z) = S(-Z) + Z.    
\end{equation*}
This implies, first of all, that the constant Sez may be taken to be --SEZ, and more importantly, that the table implementing $S$ need only be indexed from 0 to SEZ (instead of from Sez to SEZ). The machine verification of theorems relating to these issues follows in a later section. They are presented above to motivate our choice of axioms.

We postulate the existence of an integer constant, SEZ, and a table, ST, with integer values and integer indices, that satisfy the following:
\begin{equation}
    0 < \mathrm{SEZ}
\end{equation}
\begin{equation}
\bigg(\frac{P}{Q}\bigg)^{\mathrm{SEZ}+1} +1 <    
\bigg(\frac{P}{Q}\bigg)^{\mathrm{SEZ}+2}    
\end{equation}
Axiom (3) ensures that SEZ is defined so that 
$S(\mathrm{SEZ} + 1) = \lfloor s_b(\mathrm{SEZ} + 1)\rfloor = \lfloor\log_b(b^{\mathrm{SEZ}} + 1)\rfloor = \mathrm{SEZ}+ 1$. 
This is a fact that can be used in an inductive proof that for all integers 
$Z > \mathrm{SEZ}, S(Z) = Z$.

The remaining axioms deal with the nature of the finite table, ST, used to implement addition. ST is a table of nonnegative integers, indexed by an integer between zero and the essential zero. For $0\leq Z \leq \mathrm{SEZ}$,
\begin{equation}
\mathrm{ST(Z)} > 0    
\end{equation}
and
\begin{equation}
\bigg(\frac{P}{Q}\bigg)^{\mathrm{ST}(Z)} \leq    
\bigg(\frac{P}{Q}\bigg)^{Z} +1  <    
\bigg(\frac{P}{Q}\bigg)^{\mathrm{ST}(Z)+1}      
\end{equation}
Axiom (5) states that for $0\leq Z \leq \mathrm{SEZ}$, $\mathrm{ST}(Z) = \lfloor\log_b(b^Z+1)\rfloor = \lfloor s_b(Z)\rfloor$.

\section{Mechanical Verification}
The Boyer-Moore theorem prover, NQTHM, is a computer program that can be used to prove theorems stated in the computational logic of Boyer and Moore. The logic is a quantifier-free, first-order logic with equality resembling Pure LISP. Both the formal logic and the theorem prover are fully described in [5]. An interactive interface to the prover, fully described in [13], was used during the work reported here. NQTHM was used to show (a) the axioms are satisfiable for a specific choice of the base $b=3/2$, (b) the axioms are satisfiable for any choice of a base $b=P/Q$ with $1< Q< P < 2\cdot Q$, (c) multiplication and division implementations are correct for rational values that have exact representations as base $b$ logarithms, (d) addition implementation is correct in the sense that it works as well as can be expected for rational values with exact representations as base $b$ logarithms. (e) when the implementation of these three operations are used repeatedly in a program, the relative error increases in a way that can be described a priori.

In order to explore the various issues that arise when logarithmic arithmetic is formalized, we have chosen to break the implementation into several different levels, Level 1, which is described in this paper, only deals with representing arbitrarily large positive rationals. Zero, underflow and overflow can never occur in Level 1, nor is subtraction a permissible operation. The implementations of multiplication and division in Level 1 seem trivial. Although somewhat more involved, addition can be succinctly expressed using the $S(Z)$ function described above. However, just proving that these three implementations are correct in terms of nonnegative integers requires about 6,000 lines of lemmas be derived and keyed into NQTHM to guide its reasoning.

Although a library of theorems about rational numbers is available for NQTHM [56], we decided to use nonnegative integers for Level 1 of our formalization. Although this makes our axioms and theorems more intricate than they would have been if rationals had been used, this was our only option when this work began since the rationals library requires more resources than were available to us on the DEC Station R5000 that we have for running NQTHM. Although we miss the notational convenience of a predefined rational type, NQTHM is able to reason about inequalities involving nonnegative integers.

Most practical applications have used the sign/logarithm number system [30], where there is a sign bit in the representation, allowing both positive and negative rationals to be represented. It is our goal to include a sign bit in a future level of our formalization. We hope that the inclusion of a sign bit will no more than double the number of lemmas, and that these lemmas will not be significantly different from the ones we have proven for Level 1.

All the axioms from the previous section were manually translated into equivalent statements (in the LISP-like syntax of NQTHM) about nonnegative integers. These axioms were submitted to the theorem prover, and it verified that they are consistent by showing that the following choices for the base, $b= P/Q$, the constant SEZ, and the table, ST, satisfy the axioms:
\begin{equation*}
    b=\frac{3}{2}
\end{equation*}
\begin{equation*}
    \mathrm{SEZ}=1
\end{equation*}
    \begin{center}
        \begin{tabular}{|c|c|}
        \hline
        $Z$&ST$(Z)$ \\
        \hline
        0&1 \\
        1&2 \\
        \hline
        \end{tabular}
    \end{center}

Most of the axioms and lemmas involved in formalizing Level 1 logarithmic arithmetic are simple inequality predicates involving positive rational numbers, A positive rational must be translated into a pair of positive integers before submission to the theorem prover. The following definitions are based on the assumption that TOP, BOT, BASE\_TOP, BASE\_BOT are positive integers. The exponent, $E$, may be negative as well as zero or positive, and is usually a logarithmic representation. BASE\_TOP and BASE\_BOT together define a base, usually $b$. TOP and BOT together define an arbitrary finite positive rational number which is to be compared against the value $b^E$, One of the predicates needed for our axioms is
\begin{equation*}
\mathrm{L\_LESSP(TOP,BOT,BASE\_TOP,BASE\_BOT,E)}=
\Bigg(       \frac{\mathrm{TOP}}      {\mathrm{BOT}}  <
\bigg( \frac{\mathrm{BASE\_\ TOP}}{\mathrm{BASE\_\ BOT}} \bigg)^E 
\Bigg)   
\end{equation*}
which can be translated into equivalent nonnegative integer inequalities with the following definition:
\begin{verbatim}
(DEFN
  L-LESSP ( TOP BOT BASE-TOP BASE-BOT E ) 
  (IF (NEGATIVEP E)
      (LESSP (TIMES TOP
                    (EXP BASE-TOP (NEGATIVE-GUTS E)))
             (TIMES BOT
                    (EXP BASE-BOT (NEGATIVE-GUTS E))))
      (LESSP (TIMES TOP
                    (EXP BASE-BOT E))
             (TIMES BOT
                    (EXP BASE-TOP E)))))
\end{verbatim}
where {\tt (NEGATIVE-GUTS E)} is a NQTHM function that strips the sign from its negative input, $E$, leaving $|E|$. The notation {\tt L-LESSP} indicates this is the less than predicate with the arbitrary rational, TOP/BOT, on the {\it left}. {\tt L-GEQ} is the negation of {\tt L-LESSP}. Similar definitions are used for {\tt R-LESSP} and {\tt R-GEQ}, which have the arbitrary rational on the {\it right.}
For example, here are axioms (4) and (5) in NQTHM syntax:
\begin{verbatim}
  (NUMBERP (ST Z)) : S-TABLE AXIOMS
  (IMPLIES (AND (NUMBERP Z)
                (NOT (LESSP (SEZ) Z))) 
           (L-GEQ (PLUS (EXP (P) Z)
                        (EXP (Q) Z)) 
                  (EXP (Q) Z)
                  (P) (Q) (ST Z))) 
  (IMPLIES (AND (NUMBERP Z)
                (NOT (LESSP (SEZ) Z))) 
           (L-LESSP (PLUS (EXP (P) Z)
                          (EXP (Q) Z)) 
                  (EXP (Q) Z)
                  (P) (Q) ADD1 (ST Z)))
\end{verbatim}
Note that {\tt (SEZ)}, {\tt (P)} and {\tt (Q)} are argumentless functions that return the constants SEZ, $P$ and $Q$ respectively.

\section{Floor and Ceiling of the Logarithm} 

When $b = P/Q$, $0< D$, $0 <Q$, and $Q < P$, the function
\begin{equation}
    \mathrm{FLOOR\_LOG}(N,D,P,Q) = \lfloor\log_b(N/D)\rfloor,
\end{equation} 
is used to show that the axioms constrained above can be satisfied for any choice of $P$ and $Q$ that satisfies axiom (1). This function can also be used to convert rationals, which are given as pairs $(N,D)$ of nonnegative integers, into logarithmic form.
The base, represented by the pair $(P,Q)$, is assumed to be a rational larger than 1. This function is defined by cases, depending whether the positive rational represented by the pair $(N, D)$ is less than one (which is obtained by computing {\tt (CEILING-LOG>=1 D N P Q)} and negating this result) or whether the rational is greater than or equal to one (which is computed by {\tt FLOOR-LOG>=1}).
The function {\tt CEILING-LOG>=1} is designed to correctly compute $\lceil \log_{P/Q}(N/D) \rceil$ when $(N,D)$ represents a rational at least as large as 1 and $(P, Q)$ represents a rational base larger than 1. The function searches through values of the form $(P/Q)^L$, for $L$= 0,1,2,..., until $N/D < (P/Q)^L$. When the first such value is found, the function returns the value of L.   NQTHM proved that the search must end with $L < N \cdot Q$. Here is the definition as presented to the theorem prover:
\begin{verbatim}
(DEFN
   CEILING-LOG>=1-LOOP (N D P Q L I J C)
   (IF (ZEROP C) 
       (FIX L)
       (IF (LESSP (TIMES D I)
                  (TIMES W J)) 
           (CEILING-LOG>=1-LOOP N D P Q
                               (ADD1 L) 
                               (TIMES I P) 
                               (TIMES J Q) 
                               (SUB1 C))
           (FIX L))))
(DEFN
   CEILING-LOG>=1 (N D P Q)
   (CEILING-LOG>=1-LOOP N D P Q 0 1 1 (TIMES N Q)))
\end{verbatim}
{\tt CEILING-LOG>=1} initializes $L$ to 0, both $I$ and $J$ to 1, and $C$ to $N\cdot Q$, and then enters {\tt CEILING-LOG>=1-LOOP} which accumulates the values $(P/Q)^L$ in $I/J$ and limits the search to $L < N\cdot Q$. The above function returns {\tt (FIX L)} in both termination cases in order to ensure that the result is a nonnegative integer.
A similar function computes {\tt FLOOR-LOG>=1}. These definitions were presented to the theorem prover, and were shown to be correct. That is, when $0<D$, $0<N$, $0<Q$, and $Q<P$,
\begin{equation*}
\bigg(\frac{P}{Q}\bigg)^{\mathrm{FLOOR\_LOG}(N,D,P,Q)} \leq    
\frac{N}{D} <    
\bigg(\frac{P}{Q}\bigg)^{\mathrm{FLOOR\_LOG}(N,D,P,Q)+1} \end{equation*}

We want to show that our axioms can be satisfied for any choice of $P$ and $Q$ such that $0 < Q < P$, so that we can control the choice of precision. For example, $P$ and $Q$ can be chosen so that the precision is the same as for single precision floating point, We want to ensure that the axioms are not merely satisfiable for a particular base, but that the axioms also hold for useful choices of $b$. To do this, we use parameterized definitions for the constant SEZ and the table ST using {\tt FLOOR\_LOG}. The algorithms for computing the {\tt FLOOR\_LOG} are not practical to actually carry out the computations—they merely demonstrate that the axioms apply to systems with practical precisions. Given parameters $P$ and $Q$ such that $0< Q < P$, parameterized definitions that satisfy the axioms constrained earlier were defined for NQTHM:
\begin{equation*}
\mathrm{SEZ\_PQ}(P,Q) = \bigg\lfloor\log_{P/Q}\bigg(\frac{Q}{P-Q}\bigg)\bigg\rfloor
\end{equation*}
\begin{equation*}
\mathrm{ST\_PQ}(Z,P,Q) = \bigg\lfloor\log_{P/Q}\bigg(\frac{P^Z+Q^Z}{Q^Z}\bigg)\bigg\rfloor = \lfloor s_b(Z)\rfloor
\end{equation*}
Here are the axioms in terms of parameterized SEQ\_PQ and ST\_PQ
\begin{equation}
    0 < \mathrm{SEZ\_PQ}(P_1,Q_1)
\end{equation}
\begin{equation}
\bigg(\frac{P_1}{Q_1}\bigg)^{\mathrm{SEZ\_PQ}(P_1,Q_1)+1} +1 <    
\bigg(\frac{P_1}{Q_1}\bigg)^{\mathrm{SEZ\_PQ}(P_1,Q_1)+2}    
\end{equation}
For $0 \leq Z \leq \mathrm{SEZ\_PQ}(P_1,Q_1)$,
\begin{equation}
\mathrm{ST\_PQ}(Z,P_1,Q_1) \geq 0    
\end{equation}
and
\begin{equation}
\bigg(\frac{P_1}{Q_1}\bigg)^{\mathrm{ST\_PQ}(Z,P_1,Q_1)} \leq    
\bigg(\frac{P_1}{Q_1}\bigg)^{Z} +1  <    
\bigg(\frac{P_1}{Q_1}\bigg)^{\mathrm{ST\_PQ}(Z,P_1,Q_1)+1}      
\end{equation}
These were translated into nonnegative integer arithmetic, and NQTHM verified that the parameterized versions of SEZ and ST actually satisfy the axioms, as they are stored in the data base of the theorem prover, given only that the parameters $P_1$ and $Q_1$ satisfy axiom $1 < Q_1 < P_1 < 2\cdot Q_1$.

For example, the axioms can be satisfied when the base is chosen to 
be:
\begin{equation*}
    b=\frac{12,500,001}{12,500,000}.
\end{equation*}
For this choice of base, the precision is
$$\lfloor\log_2(\log_b(2))\rfloor = 23,$$
which is the precision (not counting the hidden bit) of single precision floating point systems. The number of entries in the ST table is
\begin{equation*}
\bigg\lfloor\log_b\bigg(\frac{Q}{P-Q}\bigg)\bigg\rfloor+1
=\lfloor \log_b(12,500,000)\rfloor +1 
= 204,265,491+1 \approx 2^{27.61}.
\end{equation*}

A table of this size is too large to be practical. Furthermore, the time required to compute the complete ST table using {\tt FLOOR-LOG} as defined in our formalization is astronomical because the accumulated product, {\tt I}, grows to be huge ($Q^{S(Z)}$), Since $Z > 0$, $\lfloor\log_b(2) < S(Z)\rfloor$, and so it will take at least as long to compute $S(Z)$ as it takes to compute $Q^{\lfloor \log_b(2)\rfloor}$, which in turn takes at least as long as computing the final {\tt TIMES}.
For the choice of $b$ given above that makes $F$ = 23, $\lfloor\log_b(2)\rfloor$ = 8,664,340, and so $8,664,340\cdot(12,500,000^{8,664,340})$ is the final {\tt TIMES}. Assuming a straightforward multiplication algorithm, this requires at least 
$F = \lfloor\log_e(\log_b(2))\rfloor = 23$ 
additions of $k$ bit numbers, where 
$k = 8,664,339 \cdot \lceil \log_2(12,500,000)\rceil = 199,279,820$.
This means $k \cdot F = 4,583,435,860$ bit operations occur in this final {\tt TIMES}, which is a lower bound on the number of bit operations required to compute one $S(Z)$ for any $Z > 0$ with this base using {\tt FLOOR-LOG}. The actual number of bit operations would be much greater. There are SEZ elements in the ST table, and so a lower bound on the number of bit operations to do the
final TIMES in filling each element of the ST table is 
57,292,948,250,000,000. A similar {\tt TIMES} occurs in computing $P^S(Z)$, Assuming a machine with 1 ns. per operation, a lower bound for how long it takes to fill the ST table using {\tt FLOOR-LOG} is at least two years. It would actually take much longer because of all of the other {\tt TIMES} that were ignored above.
Improvements in the implementation and efficiency of the functions that compute the floor of the logarithm are possible. For example, the techniques of [17] would allow the complete ST table to be initialized in less than two seconds. {\tt FLOOR-LOG} can serve as the specification for such faster but equivalent function.

\section{Level 1 Exact Representation}

In our formalization, we make a distinction between the representation and the (rational) value being represented. The representation $Z$ (which is an integer) exactly represents the positive rational value $N/D$ if and only if
$$\bigg(\frac{N}{D}\bigg) =
\bigg(\frac{P}{Q}\bigg)^{Z}.$$
There are two cases to consider depending on whether $Z$ is negative or nonnegative, because {\tt (EXP X E)} is only valid for a nonnegative integer, $E$. Here is what is meant by the rational $\mathrm{TOP}/\mathrm{BOT}$ being exactly representable as $Z$ in base $\mathrm{BASE\_TOP}/\mathrm{BASE\_BOT}$:
\begin{verbatim}
(DEFN
  EXACT-REP ( E TOP BOT BASE-TOP BASE-BOT ) 
  (IF (NEGATIVEP E)
    (EQUAL (TIMES TOP (EXP BASE-TOP (NEGATIVE-GUTS E))) 
           (TIMES BOT (EXP BASE-BOT (NEGATIVE-GUTS E))))
    (EQUAL (TIMES TOP (EXP BASE-BOT E)) 
           (TIMES BOT (EXP BASE-TOP E)))))
\end{verbatim}
In Level 1, the base is always $P/Q$, and the constants $P$ and $Q$ are nonnegative integers constrained to satisfy axiom (1), and so
$$\mathrm{EXACT\_REP\_LEVEL\_1}(Z, N, D) 
= \mathrm{EXACT\_REP}(Z, N, D, P,Q).$$

\section{Level 1 Multiplication and Division}

Here are the implementations in Level 1 of multiplication and division:
\begin{verbatim}
(DEFN
  MULT-LEVEL-1 (X Y) 
  (iPLUS X Y))
(DEFN
  DIV-LEVEL-1 (X Y)
  (iDIFFERENCE X Y))
\end{verbatim}
{\tt iPLUS} and {\tt iDIFFERENCE} are integer versions of addition and subtraction required by NQTHM since PLUS and DIFFERENCE only work, as expected, with nonnegative integers.
NQTHM verified that these implementations are correct. That is, if $X$ exactly represents $N_X/D_X$ and $Y$ exactly represents $N_Y/D_Y$, then $\mathrm{MULT\_LEVEL\_1(X,Y)}$
exactly represents
$$\frac{N_X}{D_X}\cdot\frac{N_Y}{D_Y}=\frac{N_X\cdot N_Y}{D_X \cdot D_Y}$$
The theorem prover verified this lemma after it was translated into its syntax:
\begin{verbatim}
(IMPLIES (AND (LESSP 0 NX) (LESSP 0 NY) 
              (LESSP 0 DX) (LESSP 0 DY)
              (EXACT-REP-LEVEL-1 X NX DX)
              (EXACT-REP-LEVEL-1 Y NY DY})
  (EXACT-REP-LEVEL-1 (MULT-LEVEL-1 X Y) (TIMES NX NY) (TIMES DY DX)))
\end{verbatim}
Also, if $X$ exactly represents $N_X/D_X$ and $Y$ exactly represents $N_Y/D_Y$ then $\mathrm{DIV\_LEVEL\_1}(X,Y)$ exactly represents
$$\frac{N_X}{D_X} / \frac{N_Y}{D_Y}=\frac{N_X\cdot D_Y}{D_X \cdot N_Y}$$
This was verified in a way similar to multiplication.

\section{Definition of S}
Since Level 1 uses an unbounded integer, $Z$, to represent arbitrarily large rational values, an infinite table would be required to store the function,
$$S(Z) = \lfloor s_b(Z)\rfloor,$$
where $b = P/Q$. This would obviously make implementation on a finite machine impossible. To overcome this, a finite table of values, ST, described by the axioms earlier, is used to define S as:
\begin{eqnarray*}
 S(Z) =   \left\{
          \begin{array}{ll} 
              0        & \mathrm{if} \quad Z < -\mathrm{SEZ} \\ 
              Z+ST(-Z) & \mathrm{if} \quad -\mathrm{SEZ} \leq Z <0 \\
              Z+ST(Z) & \mathrm{if} \quad 0 \leq Z \leq \mathrm{SEZ} \\
              Z        & \mathrm{if} \quad \mathrm{SEZ} < Z \\
          \end{array}
          \right.
\end{eqnarray*}
When $|Z|$ is larger than SEZ, the ST table is not used. For positive $Z$ close to zero, ST is used directly. For negative $Z$ close to zero, a simple computation using ST avoids having to store a table twice as large.
This definition was translated and NQTHM proved it is correct by showing that the following inequalities hold for all integers $Z$:
\begin{equation}
\bigg(\frac{P}{Q}\bigg)^{S(Z)} \leq    
\bigg(\frac{P}{Q}\bigg)^{Z} +1  <    
\bigg(\frac{P}{Q}\bigg)^{S(Z)+1}      
\end{equation}
In other words, the ratio of the rational $b^Z+ 1$ and the rational represented by $S(Z)$ is between 1 and $b$. This is as good a result as can be expected by any logarithmic (or floating point) number system with relative precision F. Indeed, many formal models [36, 9] of floating point use a constant equivalent to $b$ to describe such inherent relative error.

Proving this with only the non-negative integer library requires separate consideration of $Z < 0$ and $Z \geq 0$ for both inequalities. The details were quite involved, and are omitted.

\section{Level 1 Addition}
In Level 1, the sum of two values is approximated by one of four cases. Which case is used depends on the relative magnitude of the two values (i.e., their ratio). Since we are dealing with logarithms, the case is selected based on the difference of the representations:
\begin{eqnarray*}
    \mathrm{ADD\_LEVEL\_1}(X,Y) =   
          \left\{
          \begin{array}{ll} 
              Y        & \mathrm{if} \quad X-Y < -\mathrm{SEZ} \\ 
              X+ST(-(X-Y)) & \mathrm{if} \quad -\mathrm{SEZ} \leq X-Y <0 \\
              Y+ST(X-Y) & \mathrm{if} \quad 0 \leq X-Y \leq \mathrm{SEZ} \\
              X        & \mathrm{if} \quad \mathrm{SEZ} < X-Y \\
          \end{array}
          \right.
\end{eqnarray*}
The following theorem shows that this implementation of addition is correct in terms of the function $S$:
\begin{verbatim}
   (EQUAL (ADD-LEVEL-1 X Y) (iPLUS Y (S (iDIFFERENCE X Y))))    
\end{verbatim}
This theorem is used to assist the theorem prover in the next proof.
The final Level 1 theorems show that the implementation of addition is
correct in the sense that the following inequalities hold: If $X$ exactly represents $N_X/D_X$ and $Y$ exactly represents $N_Y/D_Y$, with $D_Y$ and $D_Y$ both positive, then
\begin{eqnarray*}
\bigg(\frac{P}{Q}\bigg)^{\mathrm{ADD\_LEVEL\_1}(X,Y)} 
& \leq & \frac{N_X}{D_X}+\frac{N_Y}{D_Y} \\
& = &\frac{N_X\cdot D_Y + N_Y\cdot D_X}{D_X \cdot D_Y} \\
& < &\bigg(\frac{P}{Q}\bigg)^{\mathrm{ADD\_LEVEL\_1}(X,Y)+1} \\     
\end{eqnarray*}
That is,
$$
\mathrm{ADD\_LEVEL\_1}(X,Y) = \bigg\lfloor 
\log_{P/Q}\bigg(\frac{N_X}{D_X} + \frac{N_Y}{D_Y} \bigg) 
\bigg\rfloor.$$

\section{Level 1 tolerances}
Using the theorems given so far to prove the correctness of Level 1 programs that use repeated multiplication and division should pose no problem since such representations will remain exact if the initial inputs are exact. However, such proofs are trivial since they simply amount to saying a program that computes sums and differences of integers is correct.
As contrasted with multiplication and division of exact values, Level 1 addition of exact values does not produce an exact result, This inherent truncation is one major obstacle in proving “floating point” programs are correct. In a program that uses Level 1 logarithmic arithmetic, the two possible sources of truncation are: (a) conversion of rational inputs to Level 1 representations, using {\tt FLOOR-LOG} and (b) addition of Level 1 representations. {\tt FLOOR-LOG} may
occasionally be exact, for example when converting the number one, but in this formalization we must assume {\tt ADD-LEVEL-1} is always inexact, even when its inputs are exact. This poses a problem in proving the correctness of programs that perform more than one addition. Furthermore, the results from such program fragments may be input into other functions that do multiplication, and so the assumption of exact input(s) will no longer hold. We need a notation to account for the accumulated error that results from repeated finite precision computation.
Past attempts to formalize this have viewed floating point values as a subset of the rationals, as they most certainly are [36, 18, 9]. On the other hand, verification of the floating point implementations [4] have concentrated on the manipulation of the representations, as we have just done in the previous sections. What is needed is a notation that merges the concepts of value and representation. This notation also must be flexible enough to describe accumulated error. This new notation, although somewhat similar to interval arithmetic [21], is specialized to the task of automated verification of logarithmic arithmetic.

We define a predicate, which we refer to as a tolerance, to state,
\begin{equation}
    TOL(P,Q,T_L,T_R,Z,N,D)= \Bigg(
    \bigg(\frac{P}{Q}\bigg)^{Z+T_L} \leq    
    \frac{N}{D}  <    
    \bigg(\frac{P}{Q}\bigg)^{Z+T_R} \Bigg)      
\end{equation}
which means a particular arbitrary rational value (defined by $N$ and $D$) and a certain machine representation ($Z$) in a certain base (defined by $P$ and $Q$) are related to one another by at most a certain kind of accumulated relative error bound (defined by $T_L$, and $T_R$). In the syntax of the theorem prover,
\begin{verbatim}
 (DEFN
    TOL (P Q TL TH Z N D)
    (AND
         (L-GEQ N D P Q (iPLUS Z TL))
         (R-GEQ N D P Q (iPLUS Z TH))))
\end{verbatim}
In Level 1, all the tolerances use base $b$, and so
$$\mathrm{TOL\_LEVEL\_1}(T_L,T_H,Z,N,D) = 
\mathrm{TOL}(P,Q,T_L,T_H,Z,N,D)$$
In essence, the tolerance notation generalizes the kind of inequalities commonly used in our axioms and theorems, such as axiom (5), except that the tolerance does not use a strict inequality. The reason for using {\tt R-GEQ}, instead of {\tt L-LESSP} is so that
\begin{verbatim}
  (EXACT-REP-LEVEL-1 Z N D)     
\end{verbatim}
means the same as
\begin{verbatim}
    (TOL-LEVEL-1 0 0 Z N D).
\end{verbatim}
Many of the axioms and theorems can be restated using this notation, for example
\begin{verbatim}
 (IMPLIES
   (AND (NUMBERP Z)
        (NOT (LESSP (SEZ) 2)))
   (TOL-LEVEL-1 0 1 (ST Z) (PLUS (EXP (P) Z) (EXP (Q) Z)) (EXP (Q) Z)))
\end{verbatim}
is similar to axiom (5) with “$\leq$" used in the right inequality instead of “$<$". With this notation, we have proven more general theorems about accumulated error. For example, NQTHM proved that the tolerances add when the representations of two positive rational values are multiplied using {\tt MULT-LEVEL-1}:
\begin{verbatim}
  (IMPLIES
     (AND (TOL-LEVEL-1 TLX THX X NX DX)
          (TOL-LEVEL-1 TLY THY Y NY DY)) 
     (TOL-LEVEL-1 (iPLUS TLX TLY) (iPLUS THX THY)
                  (MULT-LEVEL-1 X Y)
                  (TIMES NX NY) (TIMES DX DY)))
\end{verbatim}
Here, $T_{LX}$ and $T_{HX}$ define the tolerance for $X$, and $T_{LY}$ and $T_{HY}$ define the tolerance for Y, This generalizes the result given earlier only for exact inputs ($T_{LX}=T_{HX}=T_{LY}=T_{HY}=0$), and is very useful in proving later theorems.

NOTHM also proved that computing a reciprocal causes the high and low tolerances of the representation to be both negated and interchanged in the result. By combining this fact with the multiplication theorem above, NQTHM proved what happens to the tolerance of a quotient. If it is assumed $T_{LX}$, $T_{HX}$, $T_{LY}$ and $T_{HY}$ are defined as in the multiplication tolerance theorem above, NQTHM proved the low and high tolerances of {\tt (DIV-LEVEL-1 X Y)} are $T_{LX} - T_{HY}$ and $T_{HX} - T_{LY}$, respectively.

\section{Addition tolerances}

Earlier, it was shown that for exact inputs, $X$ and $Y$, $\mathrm{ADD\_LEVEL\_1}(X, Y)$ produces a representation of $b^X + b^Y$ which is as good as is possible. With the tolerance notation, the relative error bounds of this result for exact $X$ and $Y$ can be defined as $T_{LA} = 0$ and $T_{HA} = 1$. To be useful in complicated programs, $T_{LA}$ and $T_{HA}$ need to be generalized to consider the case of an inexact input.
If we assume $X$ represents $N_X/D_X$ with tolerance defined by $T_{LX}$ and $T_{HX}$ and $Y$ represents $N_Y/D_Y$ with tolerance defined by $T_{LY}$ and $T_{HY}$, then the following {\it real} inequalities hold:
\begin{equation}
\bigg(\frac{P}{Q}\bigg)^{X+T_\mathrm{LX}} \leq    
\frac{N_X}{D_X}  <    
\bigg(\frac{P}{Q}\bigg)^{X+T_\mathrm{HX}}      
\end{equation}
\begin{equation}
\bigg(\frac{P}{Q}\bigg)^{Y+T_\mathrm{LY}} \leq    
\frac{N_Y}{D_Y}  <    
\bigg(\frac{P}{Q}\bigg)^{Y+T_\mathrm{HY}}      
\end{equation}
\begin{equation}
\bigg(\frac{P}{Q}\bigg)^{X+T_\mathrm{LX}} + \bigg(\frac{P}{Q}\bigg)^{Y+T_\mathrm{LY}}\leq    
\frac{N_X}{D_X} + \frac{N_X}{D_X} <    
\bigg(\frac{P}{Q}\bigg)^{X+T_\mathrm{HX}} + \bigg(\frac{P}{Q}\bigg)^{Y+T_\mathrm{HY}}     
\end{equation}
\begin{equation}
b^{X+T_\mathrm{LX} + s_b(Y+T_\mathrm{LY}-X-T_\mathrm{LX})} \leq    
\frac{N_X}{D_X} + \frac{N_X}{D_X} <    
b^{X+T_\mathrm{HX} + s_b(Y+T_\mathrm{HY}-X-T_\mathrm{HX})}
\end{equation}
\begin{equation}
\bigg(\frac{P}{Q}\bigg)^{X + S(Y-X)+T_\mathrm{LA}} \leq    
\frac{N_X}{D_X} + \frac{N_X}{D_X} <    
\bigg(\frac{P}{Q}\bigg)^{X + S(Y-X)+T_\mathrm{HA}}
\end{equation}
where $s_b$ is the real valued addition logarithm with rational base $b= P/Q$, and the integers $T_{LA}$, and $T_{HA}$ are chosen so as to make the above inequalities true:
$$T_{LA}=T_{LX}+S(Y+T_{LY}-X-T_{LX})-S(Y-X)\leq T_{LX}+s_b(Y+T_{LY}-X-T_{LX})-S(Y-X)$$
and
$$T_{HA}=T_{HX}+S(Y+T_{HY}-X-T_{HX})-S(Y-X)+1\geq T_{HX}+s_b(Y+T_{HY}-X-T_{HX})-S(Y-X).$$

We proved the validity of these definitions for $T_{LA}$ and $T_{HA}$ with NQTHM. The proof was quite involved because of the need to translate to nonnegative integers. It required several cases, depending on the sign of $Y-X$, on whether $X+T_{LX}$ and $X+T_{HX}$ are both positive, both negative, or of mixed signs, and also on the signs of $Y+T_{LY}$ and $Y+T_{HY}$.

$X$ and $Y$ may not be known until run-time, but the tolerances typically are known during the proof of the correctness of the program. For those circumstances where no additional knowledge is available about $X$ and $Y$ during the proof, a correct, but less tight, bound on $_{LA}$ and $T_{HA}$ can be obtained. Although NQTHM does not prove it this way, we know for all real $z$ that $0 < s'(z) < 1$, and so it is reasonable to expect that similar bounds exist for the first difference of $S(Z)$ for any integer $Z$. In fact, the following theorems hold for all integers $Z$:
$$0\geq \Delta \implies 0 \leq S(Z+\Delta) -S(Z) \leq \Delta $$
and
$$0\leq \Delta \implies \Delta \leq S(Z+\Delta) -S(Z) \leq 0.$$
From this knowledge, we can determine a much simplier expression for the tolerance than $T_{LA}$ and $T_{HA}$
above. 
Let $\Delta = T_{LY}-T_{LX}$ and $Z = Y - X$, and so when $T_{LX} < T_{LY}$, we know that $\Delta > 0$, and so it is valid to define $\bar{T}_{LA} = T_{LX} +0 < T_{LA}$ in this case. Also, when $T_{LX}\geq T_{LY}$, we know $\Delta \leq 0$, and so it is correct to use $\bar{T}_{LA} = T_{LX} +\Delta < T_{LA}$. Therefore, when there is no other information about $X$ and $Y$, it is valid to use
$$\bar{T}_{LA} = \min(T_{LX},T_{LY})$$
and by similar reasoning,
$$\bar{T}_{HA} = \max(T_{HX},T_{HY}+1).$$

Although these are not as tight as would be possible if something were known about $Z$, they are good enough to mechanically verify proofs that programs involving summation of terms, such as evaluating a polynomial, have reasonable tolerances that can be determined at proof time. Because the tolerance of the accumulated sum soon grows to be larger than the tolerance of any term, most {\tt ADD\_LEVEL\_1} operations simply increment $T_{HA}$ by one. This corresponds to our intuition about what happens to accumulated relative error as a result of multiple table look ups.

\section{Taylor series example}
Polynomials derived from truncated series are often used to approximate functions. For example,
$$f(x)=1+x+\frac{x^2}{2!}+\frac{x^3}{3!}$$
provides an approximation to $e^x$ for small positive $x$. A Level 1 program implementing this computation,
\begin{verbatim}
(DEFN F (X) 
  (ADD-LEVEL-1
    (DIV-LEVEL-1 (MULT-LEVEL-1 X (MULT-LEVEL-1 X X)) 
                 (FLOOR-LOG 6 1 (P) (Q)))
    (ADD-LEVEL-1
       (DIV-LEVEL-1 (MULT-LEVEL-1 X X)
                    (FLOOR-LOG 2 1 (P) (Q))) 
       (ADD-LEVEL-1 X 0))))
\end{verbatim}
was submitted to NQTHM together with several lemmas about the tolerances of each subexpression. NQTHM proved that if $T_{LX} = 0$ and $T_{HX}= 1$ (as would be true if $x$ had been converted with FLOOR-LOG), then {\tt (F X)} approximates $f(x)$ with a tolerance of 
$T_{FL} = -1$ and $T_{FH} = 4$. In other words, the logarithmic representation of $f(z)$, $\lfloor\log_b(f(x))\rfloor$, is one of the six integers $F(X)-1$, $F(X)$, $F(X) + 1$, $F(X) +2$, $F(X)+3$ and $F(X) +4$. That is, the value of $F(X)$, $b^{F(X)}$, is one of the six representable values $b^{\lfloor\log_b(f(x))\rfloor}\cdot b^{-4}$,
$b^{\lfloor\log_b(f(x))\rfloor}\cdot b^{-3}$,
$b^{\lfloor\log_b(f(x))\rfloor}\cdot b^{-2}$,
$b^{\lfloor\log_b(f(x))\rfloor}\cdot b^{-1}$,
$b^{\lfloor\log_b(f(x))\rfloor}\cdot$ and
$b^{\lfloor\log_b(f(x))\rfloor}\cdot b^{1}$.

To describe the rational value, $f(x)$, being approximated here once again requires translating to statements about non-negative integers. In this case, however, the non-negative integer description is quite unwieldly. The unreduced numerator of $f(x)$ is:
\begin{verbatim}
  (PLUS (TIMES (TIMES (TIMES WX (TIMES NX NX)) 
  1) (TIMES (TIMES 2 (TIMES DK DX}) (TIMES DX 
  1))) (TIMES (PLUS (TIMES (TIMES (TIMES NX 
  NX) 1) (TIMES DX 1)) (TIMES (PLUS (TIMES NX 
  1) (TIMES 1 DX}) (TIMES 2 (TIMES DK DX)))) 
  (TIMES 6 (TIMES DX (TIMES DK DX))))).
\end{verbatim}
A programmer could determine that the above is 
$12\cdot D_X^6\cdot f(N_X/D_X)$,
but such manual manipulations are tedious and error prone. 
If we had been able to use the NQTHM rationals library for Level 1, we could have described $f(x)$ directly, and NQTHM would have translated to such an expression. As a workaround to this problem, we wrote an unverified program to generate the lemmas involving such expressions. One lemma is generated for each subexpression. We did not bother to verify this lemma generator, since its output is intended to be input. to NQTHM. This program inserts into each lemma the hints required for NOTHM to arrive at the proper conclusions, and so the generator is somewhat more than just a rational to integer translator.

It is interesting to note that addition tolerances are not associative. Had $f(x)$ been evaluated in the reverse order, the tolerances would have been $T_{FL} = -1$ and $T_{FH} = 6$.

\section{Level 2}
The representation for Level 1 is an unbounded integer, which is somewhat analagous to a member of the infinite set, $F^*$, described in the Language Compatible Arithmetic Standard (LCAS) [18]. The axioms can be made more realistic with the inclusion of the concepts of underflow and overflow, which occur when numbers are represented with a bounded exponent. The only additional axiom for Level 2 postulates two integer constants, {\tt MIN\_LEVEL\_2} and {\tt MAX\_LEVEL\_2}, such that
\begin{equation}
    \mathrm{MIN\_LEVEL\_2} < \mathrm{MAX\_LEVEL\_2}. 
\end{equation}
We define the predicate {\tt IN-RANGE-LEVEL-2} to indicate if a Level 1 argument is inside of the allowed range for Level 2:
\begin{verbatim}
(DEFN
   IN-RANGE-LEVEL-2 (X)
        (AND (NOT (iLESSP X (MIN-LEVEL-2))) 
             (NOT (iLESSP (MAX-LEVEL-2) X))))
\end{verbatim}
Also, {\tt (SIGNAL-OUT-RANGE-LEVEL-2)} returns MAX\_LEVEL\_2 + 1 which represents a result that overflowed or underflowed. With a clipping function,
\begin{verbatim}
 (DEFN
   CLIP-LEVEL-2 (X Y RESULT)
     (IF (AND (IN-RANGE-LEVEL-2 X) 
              (IN-RANGE-LEVEL-2 Y)
              (IN-RANGE-LEVEL-2 RESULT))
        RESULT
        (SIGNAL-OUT-RANGE-LEVEL-2))) 
\end{verbatim}
and definitions such as:
\begin{verbatim}
 (DEFN
   MULT-LEVEL-2 (X Y)
   (CLIP-LEVEL-2 X Y (MULT-LEVEL-1 X Y)))) 
\end{verbatim}   
we can have theorems like:
\begin{verbatim}
 (IMPLIES
    (IN-RANGE-LEVEL-2 (MULT-LEVEL-2 X Y)) 
    (AND (IN-RANGE-LEVEL-2 X)
         (IN-RANGE-LEVEL-2 Y) 
         (EQUAL (MULT-LEVEL-1 X Y)
                (MULT-LEVEL-2 X Y))))
\end{verbatim}
that state Level 1 proofs validate Level 2 programs provided that the machine running the Level 2 program never detects an underflow or overflaw. Therefore, Level 2 uses a machine with a finite word size and finite tables, and so Level 2 representations have both finite range and finite precision.

\section{Conclusions}
We have mechanically verified the first levels of what we hope to be several levels of increasingly realistic logarithmic arithmetic implementations. Although Level 1 may seem very simplistic, the proofs required to reach this stage were non- trivial. We did not present our subtraction axioms, which requires a table, 
DT(Z) = $\lceil \log_b|1 -b^Z |\rceil$, of the quantized subtraction logarithm. These axioms are similar to (2)-(5), however the DT table is indexed from 1 to SEZ+1. A complete Level 1 implementation requires verifying an implementation of the computation of the absolute value of the difference of positive rational values using DT. Although, as we have shown, it is possible to prove useful theorems about the tolerances of ADD\_LEVEL\_1 programs without knowing $Z$, this is not possible for programs that use subtraction. A proof about subtraction must have some information about $Z$ since 
$|d'(z)| \rightarrow -\infty$ as $z$ approaches zero. In other words, when $Z$ is near zero, severe cancellation occurs. Overcoming the formalization problems presented by this inherent property of floating point subtraction [20] is beyond the scope of this paper.

Levels 1 and 2 use simple table look up to approximate $S(Z)$. However, as shown earlier, it is impractical to have a complete table of the more than $2^{27}$ words required for single precision ($F = 23$), and so much more sophisticated algorithms than what we have considered here will be required. The addition logarithm function is often approximated by interpolation [37]. Therefore, to verily a practical, single precision logarithmic arithmetic implementation re- quires verifying fairly complicated ``floating point” programs which compute $S(Z)$ from a much smaller table. Initial research in this area leads us to think the divided difference method is most amenable to formalizing, but the proofs are very complicated, and we have not yet submitted any to NOTHM, We expect the tolerance notation described above to be helpful in such proofs.

The functional programming paradigm of NQTHM is quite different from the typical application of logarithmic arithmetic. More realistic models of computing devices could be developed and logarithmic arithmetic correctly implemented on them. The hope is that eventually an implementation ($F = 23$)
in a machine language that has been formally specified in NQTHM, such as MC68020 [7] or PITON [22], could be verified. Also, gate level verification [12] of logarithmic arithmetic hardware could be possible.
We are pleased with NQTHM’s ability to deal with the problems presented here. We hope to have a better computer for running NQTHM in the future, and to use the rationals library in later levels of the formalization, which should simplify the proofs.
We will provide a machine readable copy of the proofs described here ta interested parties who request it. NQTHM itself is available via anonymous ftp. The conditions for its use established by its developers and the directions for obtaining it are given in [5].


\begin{thebibliography}{999}

\bibitem{arn91}
M. Arnold,  T. Bailey,  J. Cowles and  J. Cupal,
{\em Proceedings of the  International AMSE Conference Neural Networks} San Diego, California, vol. 1, pp. 75--86,  29-31 May 1991.


\bibitem{arn92c}
M. G. Arnold,  
T. A. Bailey, J. R. Cowles and M. D. Winkel,
``Applying Features of IEEE 754 to Sign/Logarithm Arithmetic,"
{\em IEEE Transactions on Computers}, vol. 41, pp. 1040--1050, August 1992.

\bibitem{bar85} 
J. L. Barlow and E. H. Bareiss, 
``On Roundoff Error Distributions in 
Floating Point and Logarithmetic Arithmetic," {\em Computing}, vol. 34, pp. 325-347, 1985.

\bibitem{bar89}
G. Barrett, 
``Formal methods applied to a floating-point number system,” 
{\em IEEE Trans. Software Eng.}, vol. 15, pp. 611-621, May 1989.

\bibitem{boy88}
R. S. Boyer and J S. Moore. 
{\em A Computational Logic Handbook}, 
Academic Press, Boston, 1988.

\bibitem{bev89}
W.R. Bevier, 
``Kit and the short stack,” 
{\em J. Automated Reasoning}, vol. 5, pp. 519-430, 1989.


\bibitem{boy91}
B.S. Boyer and Yuan Yu, 
``Automated correctness proofs of machine code programs for a commercial microprocessor,” Tech. Report TH-91-33, Comput. Sei. and Math. Dept., Univ. Texas Austin, 1991.

\bibitem{bre73}
R. P. Brent, ``On the precision attainable with various floating point number systems,” 
{\em IEEE Trans. Comput.}, vol. C-22, pp. 601-606, June 1973.

\bibitem{bro81}
W. S. Brown, 
``A Simple but Realistic Model of Floating-Point Computation,"{\em ACM Transactions on Mathematical Software}, vol. 7, no. 4,
pp. 445-480, December 1981.

\bibitem{fastbyte}
``Fast Floating Point C Functions,” BYTE, vol. 16, no. 7, p. T2MW-8, July 1991.

\bibitem{hol80}
J. E. Holm, ``Floating-Point Arithmetic and Correctness Proofs,” Ph.D. Thesis, Cornell Univ., 1980.

\bibitem{hun89} 
W. A. Hunt,
``Microprocessor design verification,” {\em J. Automated Reasoning}, vol. 5, pp. 429-460, 1989.

\bibitem{kau88} M. Kaufmann, ``A user’s manual for an interactive enhancement to the Boyer-Moore theorem prover,” Technical Report 19, Computational Logic Inc., Austin, Texas, 1988.


\bibitem{kin71}
N. G. Kingsbury and P. J. W. Rayner,
``Digital Filtering Using Logarithmic Arithmetic,"
{\em Electronics Letters}, vol. 7, no. 2, pp. 56-58, 28 January 1971.

\bibitem{kur90}
T. Kurokawa and T. Mizukoshi,
``Fast method of geometrical picture transformation using
logarithmic number systems and its application for computer
graphics,'' {\em Proceedings of the SPIE, Visual Communications and Image Processing '90}, vol. 1360, pt. 3, pp. 1479--1490, Lausanne, Switzerland, October 1-4 1990.

\bibitem{kur91}
T. Kurokawa and T. Mizukoshi, 
``A Fast and Simple Method for Curve Drawing---A New Approach using Logarithmic Number Systems," {\em Journal of Information Processing}, vol. 14, pp. 144-152, 1991.

\bibitem{lai91}
F. Lai and C. E. Wu,
``A Hybrid Number System Processor with
Geometric and Complex Arithmetic Capabilities,''
{\em IEEE Transactions on Computers}, vol. 40, pp. 952--962, August 1991.

\bibitem{langcomp}
{\em Language Compatible Arithmetic Standard, version 3.0, X3T2/91-006}, ISOJIECITCL/SC22/WG11/N212, American National Standards Institute, Dec. 26, 1990.

\bibitem{lew85}
D. M. Lewis, 
``A Hardware Engine for Analog Mode Simulation of MOS Digital Circuits," {\em 22nd Design Automation Conference}, pp. 345-351, 1985.

\bibitem{mar73}
J. D. Marasa and D. W. Matula,
``A Simulative Study  of Correlated Error in Various Finite-Precision
Arithmetics,''
{\em IEEE Transactions on Computers}, vol. C-22, pp. 587-597, June 1973.

\bibitem{mor66}
JS. Moore, {\em Interval Analysis}, Prentice Hall, 1966.

\bibitem{mor88}
JS. Moore, PITON: A Verified Assembly Language, Tech. Report 22, Computational Logic, Inc. 1988.

\bibitem{mor89}
JS. Moore, ``A mechanically verified language implementation,” {\em J. of Automated Reasoning}, vol, 5, pp, 461-492, 1989.

\bibitem{mor90}
R. E. Morley, T. J. Sullivan and G. L. Engel,
``VLSI Based Design of a Battery-operated Hearing Aid,''{\em Southcon/90}, Orlando, Florida, pp. 55--59, 20-22 March 1990.

\bibitem{mur89}
H. Murveit, et al.,
``A large-vocabulary real-time continuous-speech recognition system,''{\em Proceedings of the IEEE International Conference on Acoustics, Speech, and
Signal Processing},
ICASSP-89. pp. 789--792, Glasgow, 1989.

\bibitem{pap88}
G. M. Papadourakis and H. Andre, 
``High Speed Implementation of Matrix Inversion Algorithms in Orthogonal Systolic Architectures,'' {\em Proceedings of the IEEE Southeastern Conference}, IEEE catalog 88CH2571-8, pp. 200--204, 1988.

\bibitem{pic89}
L. Pickett, ``Soft Co-Processors,” Log Point Systems, Mountain View, CA, 1989.

\bibitem{rum85} 
S. M. Rump, 
``Higher Order Computer Arithmetic,” {\em Proc. 7th Symp. Comput. Arithmetic}, pp. 302-308, 1985.

\bibitem{sto86}
T. Stouraitis, {\em Logarithmic Number System Theory, Analysis, and Design}, Ph.D.  Dissertation, University of Florida, Gainesville, 1986.

\bibitem{swa75} 
E. E. Swartzlander and A. G. Alexopoulos, 
``The Sign/Logarithm Number System,'' 
{\em IEEE Transactions on Computers}, vol. C-24, pp. 1238--1242, December 1975.

\bibitem{swa80} 
E. E. Swartzlander, et al., 
``Arithmetic for Ultra High Speed Tomography," 
{\em IEEE Transactions on Computers}, vol. C-29, pp. 341-353, 1980. 

\bibitem{swa83}
E. E. Swartzlander, D. Chandra, T. Nagle, and S. A. Starks,
``Sign/logarithm Arithmetic for FFT Implementation," {\em IEEE Transactions on Computers}, vol. C-32, pp. 526-534, 1983.

\bibitem{tur91}
P. J. Turner, ``Implementation and analysis of extended SLI operations,” 
{\em Proc. 10th Symp. Comput. Arithmetic}, pp. 118-126, 1991.

\bibitem{tay88} 
F. J. Taylor, R. Gill, J. Joseph and J. Radke,
``A 20 Bit Logarithmic Number System Processor,''
{\em IEEE Transactions on Computers}, vol. C-37, pp. 190--199, 1988.

\bibitem{vol92}
V. L. Volkov and P. V. Pakshin,
``The Logarithmic Number System in Control Algorithms and Information Processing,'' {\em Soviet Journal of Computer and System Sciences}, vol. 20, no. 1, pp. 132-138, 1992.

\bibitem{wil90}
M. Wilding, 
``A mechanically-checked correctness proof of a floating-point search program,” Technical Report 56, Computational Logic Inc., Austin, Texas, 1990.

\bibitem{yu90}
L. K. Yu,
{\em The Design and Implementation of a 30-bit Logarithmic
Number System Processor}, 
M.A.Sc. Thesis, University of Toronto, Canada, 1990.

\end{thebibliography}
\end{document}